\documentclass{article}
\usepackage{amsmath}

\input{tcilatex}
\begin{document}

\title{\textbf{\ 2D Schr\"{o}dinger Equation with Singular Even-Power and
Inverse-Power Potentials in Non-Commutaive Complex space }}
\date{}
\author{Slimane Zaim and Abdelkader Bahache \\
D\'{e}partement des Sciences de la Mati\'{e}re,\\
Facult\'{e} des Sciences, Universit\'{e} Hadj Lakhdar Batna, Algeria. \\
Facult\'{e} de Physique, Universit\'{e} STHB, Algeria. }
\maketitle

\begin{abstract}
We obtain exact solutions of the 2D Schr\"{o}dinger equation with the
Singular Even-Power and Inverse-Power Potentials in non-commutative complex
space, using the Power-series expansion method. Hence we can say that the
Schr\"{o}dinger equation in non-commutative complex space describes to the
particles with spin (1/2)in an external uniform magnitic field. Where the
noncommutativity play the role of magnetic field with created the total
magnetic moment of particle with spin 1/2, who in turn shifted the spectrum
of energy. Such effects are similar to the Zeeman splitting in a commutative
space.


Pacs numbers:11.10.Nx, 32.30-r, 03.65-w
\end{abstract}

\section{introduction:}

The idea of a noncommutative (NC) space is not new. It can be traced back to
Heisenberg, Pauli etc. The Heisenberg algebra:

\begin{equation}
\left[ x^{i},p_{j}\right] =i\delta ^{ij},\left[ p_{i},p_{j}\right] =0 
\tag{1}
\end{equation}

is extended with new NC commutation relations between the position
coordinate operators themselves:

\begin{equation}
\left[ \hat{x}_{i},\hat{x}_{j}\right] =i\theta _{ij}  \tag{2}
\end{equation}

This leads to new uncertainty relations:%
\begin{equation}
\Delta \hat{x}^{i}\Delta \hat{x}^{j}\succeq \frac{1}{2}\left\vert \theta
^{ij}\right\vert  \tag{3}
\end{equation}

which are a good analogy to the Heisenberg uncertainty relations. Where $%
\hat{x}^{i}=x^{i}-\frac{\theta ^{ij}}{2}p_{j}$ and the parameter $\theta
^{ij}$ is an antisymmetric real matrix of dimension length-square for the
noncommutative canonical-type space. Than the space non-commutativity is an
extension of quantum mechanics, and a cutoff could provide a solution to the
infinities appearing in quantum field theory. An example of a system where
space coordinates do not commute is that of a particle in a strong magnetic
field. There has recently been a lot of interest in the study of
noncommutative canonical-type quantum mechanics $\left[ 1,2\right] $. On
other hand the solutions of classical dynamical problems of physical systems
obtained in terms of complex space variables are well-known. There are also
interests in the complex quantum mechanic systems (in two dimentions) $\left[
1-3\right] ,$ in which we consider a quantum free particle with Singular
Even-Power and Invers-Power potentials in noncommutative complex quantum
space (so coordinate and momentum operators of this space are written $\hat{z%
}=\hat{x}+i\hat{y}$ and $p_{\hat{z}}=\left( p_{x}-ip_{y}\right) /2$) .

Furthermore in particular the Singular Even-Power and Invers-Power
potentials is useful to study the atomics physics and optical physics $%
[3-12].$In this work, we study the effect of the non-commutativity to the
free particle with \ \ Singular Even-Power and Invers-Power potentials.

This paper is organized as follows. In section $2$ we derive the deformed
Schr\"{o}dinger equation for free particle with Singular Even-Power and
Invers-Power potentials in noncommutative space, we solve this equation and
obtain the non-commutative modification of the energy levels\textbf{\ }. In
section $3$ we derive the deformed Schr\"{o}dinger equation for free
particle with Singular Even-Power and Invers-Power potentials in
noncommutative complex space, we solve this equation and obtain the
non-commutative modification of the energy levels. Finally, in section $4$,
we draw our conclusions.

\section{Schr\"{o}dinger equation with central potential in\ non-commutative
space:}

The $2D$ Schr\"{o}dinger equation with the central potential $V\left( \hat{r}%
\right) $ (Singular Even-Power or Invers-Power potentials), in
noncommutative space, which have the following form:

\begin{equation}
\left( -\frac{1}{2m}\Delta +V\left( \hat{r}\right) \right) \Psi \left(
r,\varphi \right) =\hat{E}\Psi \left( r,\varphi \right)  \tag{4}
\end{equation}

where $\hat{r}$ in noncommutative space as:

\begin{equation}
\hat{r}=r-\frac{\theta \cdot L}{2r}+\mathcal{O}\left( \Theta ^{2}\right)  
\tag{5}
\end{equation}

Then eq.$(4)$ will be written to the form:

\begin{equation}
\left( -\frac{1}{2m}\Delta +V\left( r-\frac{\theta \cdot L}{2r}\right)
\right) \Psi \left( r,\varphi \right) =\hat{E}\Psi \left( r,\varphi \right) 
\tag{6}
\end{equation}

The solution of eq. $(7)$ in polar coordinates $(r,\varphi )$ takes the
separable form:

\begin{equation}
\Psi \left( r,\varphi \right) =\frac{1}{\sqrt{2\pi }}r^{-1/2}R_{m}\left(
r\right) \exp \left( \pm im\varphi \right)   \tag{7}
\end{equation}

Then eq. $(7)$ reduces to the radial equation:

\begin{equation}
\left( -\frac{1}{2m}\left[ \frac{d^{2}}{dr^{2}}-\frac{m^{2}-1/4}{r^{2}}%
\right] +V\left( r-\frac{\theta \cdot L}{2r}\right) \right) R\left( r\right)
=\hat{E}R\left( r\right)   \tag{8}
\end{equation}

\subsection{The Singular Even-Power potential}

In the noncommutative spac the Singular Even-Power potential $V\left( r-%
\frac{\theta \cdot L}{2r}\right) $ in eq.$(8)$ takes the form:

\begin{equation}
V\left( r-\frac{\theta \cdot L}{2r}\right) =\hat{V}\left( r\right)
+V_{NC}^{\theta }\left( r\right)   \tag{9}
\end{equation}

where

\begin{equation}
\hat{V}\left( r\right) =ar^{2}+br^{-2}+\hat{c}r^{-4}+\hat{d}r^{-6}-a\theta
\cdot L+\mathcal{O}\left( \Theta ^{2}\right)   \tag{10}
\end{equation}

and $V_{NC}^{\theta }\left( r\right) $ is the perturbation term up to $%
\mathcal{O}\left( \Theta ^{2}\right) $ which takes the form:

\begin{equation}
V_{NC}^{\theta }\left( r\right) =\frac{d\theta \cdot L}{r^{8}}.  \tag{11}
\end{equation}

and

\begin{equation}
\hat{c}=c+b\theta \cdot L,\text{ }\hat{d}=d+2c\theta \cdot L  \tag{12}
\end{equation}

For simplicity, first of all, we choose the coordinate system $\left(
x,y,z\right) $ so that $\theta ^{xy}=-\theta ^{yx}=\theta \delta
^{xy}=\theta _{z},$ such that $\theta \cdot L=\theta L_{z}$ and assume that
the other components are all zero ,then the noncommutative Singular
Even-Power potential up to $\mathcal{O}\left( \Theta ^{2}\right) $ as the
form:

\begin{equation}
\hat{V}\left( r\right) +V_{NC}^{\theta }\left( r\right) =\hat{V}_{O}\left(
r\right) +V_{\text{pert}}^{\theta }  \tag{13}
\end{equation}

where

\begin{equation}
\left\{ 
\begin{array}{c}
\hat{V}_{O}\left( r\right) =-a\theta m+ar^{2}+br^{-2}+\hat{c}r^{-4}+\hat{d}%
r^{-6}\text{\ \ ,} \\ 
V_{\text{pert}}^{\theta }=V_{NC}^{\theta }\left( r\right) =\frac{d\theta m}{%
r^{8}}%
\end{array}%
\right.   \tag{14}
\end{equation}

and

\begin{equation}
\hat{c}=c+b\theta m,\text{ }\hat{d}=d+2c\theta m  \tag{15}
\end{equation}

To investigate the modification of the energy levels by eq. $(14)$, we use
the first-order perturbation theory. The spectrum of $\hat{H}_{0}=-\frac{1}{%
2m}\Delta +\hat{V}_{O}\left( r\right) ,$ and the corresponding radial wave
functions are well-known and given by $\left[ 13\right] $:

\begin{equation}
\left\{ 
\begin{array}{c}
\Phi \left( \varphi \right) =\frac{1}{\sqrt{2\pi }}\exp \left( \pm im\varphi
\right) \text{ , }m=0;1,2... \\ 
R_{m}^{n}\left( r\right) =\exp \left( -\frac{ar^{2}+\sqrt{\hat{d}}r^{-2}}{2}%
\right) \left( \underset{p=0}{\sum^{p=n}}a_{n}r^{n+\hat{\delta}}\right) 
\end{array}%
\right. ,  \tag{16}
\end{equation}

where

\begin{equation}
\hat{\delta}=3/2+\hat{c}/\left( 2\sqrt{\hat{d}}\right)   \tag{17}
\end{equation}

and the noncommutative energie eigenvalue up to $\mathcal{O}\left( \Theta
^{2}\right) $ as:

\begin{equation}
\hat{E}_{n,m}=\sqrt{a}\left( 2\hat{\mu}+4+4n\right) -\theta am,\text{ }\hat{%
\mu}=\hat{c}/\left( 2\sqrt{\hat{d}}\right)  \tag{18}
\end{equation}

the noncommutative correction of the energy levels in the first order of $%
\theta $ to the $n^{th}$ of excitation state:

\begin{equation}
\Delta E_{NC}^{\theta }=d\theta mA  \tag{19}
\end{equation}

where

\begin{equation}
A=\int \left[ r^{-4}\exp \left( -\frac{ar+br^{2}}{2\sqrt{b}}\right) \left( 
\underset{p=0}{\sum^{p=n}}a_{n}r^{n+\hat{\delta}}\right) \right] ^{2}rdr 
\tag{20}
\end{equation}

\subsection{The Invers-Power potential}

In noncommutative space the Invers-Power potential $V\left( r-\frac{\theta
\cdot L}{2r}\right) $ in eq.$(8)$ takes the form:

\begin{equation}
V\left( \hat{r}\right) =\hat{V}_{O}\left( r\right) +V_{\text{pert}}^{\theta
}\left( r\right)  \tag{21}
\end{equation}

where

\begin{equation}
\hat{V}_{O}\left( r\right) =ar^{-1}+br^{-2}+\hat{c}r^{-3}+\hat{d}r^{-4}+%
\mathcal{O}\left( \Theta ^{2}\right)  \tag{22}
\end{equation}

and $V_{NC}^{\theta }\left( r\right) $ is the perturbation term up to $%
\mathcal{O}\left( \Theta ^{2}\right) $ which takes the form:

\begin{equation}
V_{\text{pert}}^{\theta }\left( r\right) =\left( -\frac{2c}{r^{5}}-\frac{3d}{%
r^{6}}\right) \theta m  \tag{23}
\end{equation}

and

\begin{equation}
\hat{c}=c+a\theta m,\text{ }\hat{d}=d-b\theta m  \tag{24}
\end{equation}

To investigate the modification of the energy levels by eq. $(23)$, we use
the first-order perturbation theory. The spectrum of $\hat{H}_{0}=-\frac{1}{%
2m}\Delta +\hat{V}_{O}\left( r\right) ,$ and the corresponding radial wave
functions are well-known and given by $\left[ 14\right] $:

\begin{equation}
\left\{ 
\begin{array}{c}
R_{m}^{n}\left( r\right) =N_{m}r^{\hat{c}}\left(
\dprod\limits_{i=1}^{m}\left( r-\sigma _{i}^{m}\right) \right) \exp \left( 
\frac{a+br^{2}}{r}\right) , \\ 
\hat{E}_{m}=-\left( \frac{\hat{\omega}_{1}\pm \sqrt{\hat{\omega}_{1}^{2}-4%
\hat{D}\left( 2m+\hat{\mu}\right) \left( \sum_{i=1}^{m}\sigma _{i}^{m}+\sqrt{%
A}\right) }}{4\left( \sum_{i=1}^{m}\sigma _{i}^{m}+\sqrt{A}\right) }\right)
^{2}.%
\end{array}%
\right.  \tag{25}
\end{equation}

where

\begin{equation}
\hat{\omega}_{1}=\hat{\lambda}+m(2m+\hat{\mu})-m(m-1),\text{ }\hat{D}=2\text{
}\hat{b}(\hat{c}+m),A=a^{2}  \tag{26}
\end{equation}

and

\begin{equation}
\hat{\lambda}=(1+\hat{\mu})(2m+\hat{\mu})+m(m-1)+2\text{ }\hat{b}\left(
\sum_{i=1}^{m}\sigma _{i}^{m}+\sqrt{A}\right) ,\hat{\mu}=\hat{c}-1  \tag{27}
\end{equation}

the non-commutative correction of the energy levels in the first order of $%
\theta $ to the $p^{th}$ of excitation state:

\begin{equation}
\Delta E_{NC}^{\theta }=-\theta m\left( 2cf(5)+3df(6)\right)  \tag{28}
\end{equation}

where

\begin{equation}
f(5)=\int r^{-4}\left[ N_{m}r^{\hat{c}}\left( \dprod\limits_{i=1}^{m}\left(
r-\sigma _{i}^{m}\right) \right) \exp \left( \frac{a+br^{2}}{r}\right) %
\right] ^{2}dr  \tag{29}
\end{equation}

and

\begin{equation}
f(6)=\int r^{-5}\left[ N_{m}r^{\hat{c}}\left( \dprod\limits_{i=1}^{m}\left(
r-\sigma _{i}^{m}\right) \right) \exp \left( \frac{a+br^{2}}{r}\right) %
\right] ^{2}dr  \tag{30}
\end{equation}

We have shown that the energy spectrum depends on $m$. Then we deduced that
the non-commutativity plays the role of magnetic field.

\section{Schr\"{o}dinger equation with Singular Even-Power and Invers-Power
potentials in\ non-commutative complex space:}

In two dimensionals space , the complex coordinates system $\left( z,\bar{z}%
\right) $ and momentum $\left( p_{z},p_{\bar{z}}\right) $ as dened by $\left[
1,2\right] $:

\begin{equation}
\left\{ 
\begin{array}{ccc}
z=x+iy & \text{,} & \bar{z}=x-iy \\ 
& \text{and} &  \\ 
p_{z}=\frac{1}{2}\left( p_{x}-ip_{y}\right) & \text{,} & p_{\bar{z}}=-\bar{p}%
_{z}=\frac{1}{2}\left( p_{x}+ip_{y}\right)%
\end{array}%
\right.  \tag{31}
\end{equation}

We are interested to introduce the non-commutative complex operators
coordinates and their momentums in a $2D$ complex space as follows:

\begin{equation}
\left\{ 
\begin{array}{c}
\hat{z}=\hat{x}+i\hat{y}=z+i\theta p_{\bar{z}} \\ 
\widehat{\bar{z}}=\hat{x}-i\hat{y}=\bar{z}-i\theta ^{xy}p_{z} \\ 
\hat{p}_{z}=p_{z},\text{ \ \ \ }\hat{p}_{\bar{z}}=p_{\bar{z}}%
\end{array}%
\right.  \tag{32}
\end{equation}

The noncommutative algabra $(2)$ can be written as:%
\begin{equation}
\left[ \hat{z},\widehat{\bar{z}}\right] =2\theta ,\left[ \hat{z},p_{\hat{z}}%
\right] =\left[ \widehat{\bar{z}},p_{\widehat{\hat{z}}}\right] =0,\left[ 
\hat{z},p_{\widehat{\hat{z}}}\right] =\left[ \widehat{\bar{z}},p_{\hat{z}}%
\right] =2\hbar ,\ \left[ p_{\hat{z}},p_{\widehat{\hat{z}}}\right] =0 
\tag{33}
\end{equation}

Then the noncommutative complex operators coordinates as not $PT$ sym- metric

\begin{equation}
PT\hat{z}PT\neq -\widehat{\bar{z}}  \tag{34}
\end{equation}

In the noncommutative complex space we notic that $\hat{z}\widehat{\bar{z}}%
\neq \widehat{\bar{z}}\hat{z}.$Then we can show that, in the first order of
the parameter $\theta \left[ 2\right] $: 
\begin{eqnarray}
\hat{z}\widehat{\bar{z}} &=&z\bar{z}-\theta \left( L_{z}-1\right)   \notag \\
&=&z\bar{z}-\theta \left( L_{z}+2s_{z}\right) ,s_{z}=-1/2  \TCItag{35} \\
\widehat{\bar{z}}\hat{z} &=&z\bar{z}-\theta \left( L_{z}+1\right)   \notag \\
&=&z\bar{z}-\theta \left( L_{z}+2s_{z}\right) ,s_{z}=-1/2  \TCItag{36}
\end{eqnarray}

\bigskip

The momentum operator $\hat{p}^{2}$ can be written in $2D$ non-commutative
Complex space as follows:

\begin{eqnarray}
\hat{p}^{2} &=&4p_{\bar{z}}p_{z}=\left( p_{x}+ip_{y}\right) \left(
p_{x}-ip_{y}\right)   \notag \\
&=&4p_{z}p_{\bar{z}}=p^{2}  \TCItag{37}
\end{eqnarray}

In the noncommutative complex coordinate $\left( \hat{z},\overline{\hat{z}}%
\right) $ ,the noncommutative Hamiltonian associated by central potential $%
V\left( \hat{r}\right) $ as:

\begin{equation}
\hat{H}_{NC}=\left( 
\begin{array}{cc}
\frac{2}{m}p_{\hat{z}}p_{\widehat{\bar{z}}}+V\left( \hat{r}=\sqrt{\hat{z}%
\widehat{\bar{z}}}\right)  & 0 \\ 
0 & \frac{2}{m}p_{\hat{z}}p_{\widehat{\bar{z}}}+V\left( \hat{r}=\sqrt{%
\widehat{\bar{z}}\hat{z}}\right) 
\end{array}%
\right)   \tag{38}
\end{equation}

This Hamiltonian is Hermitian and represents a antiparticle withe spin $1/2.$%
The eigenfunction of the system is described by double-component spinor:

\begin{equation}
\Psi \left( z,\bar{z}\right) =\left( 
\begin{array}{c}
\Psi ^{-}\left( z,\bar{z}\right)  \\ 
\Psi ^{+}\left( z,\bar{z}\right) 
\end{array}%
\right)   \tag{39}
\end{equation}

Where the Schr\"{o}dinger equation of the system is described by two
equations:

\begin{equation}
\left( -\frac{1}{2m}\Delta +V\left( \sqrt{z\bar{z}}-\frac{\theta \left(
L_{z}+2s_{z}\right) }{2\sqrt{z\bar{z}}}\right) \right) \left( 
\begin{array}{c}
\Psi ^{-}\left( z,\bar{z}\right)  \\ 
\Psi ^{+}\left( z,\bar{z}\right) 
\end{array}%
\right) =\hat{E}\left( 
\begin{array}{c}
\Psi ^{-}\left( z,\bar{z}\right)  \\ 
\Psi ^{+}\left( z,\bar{z}\right) 
\end{array}%
\right) ,\text{ \ }s_{z}=\mp 1/2  \tag{40}
\end{equation}

where the sign $\left( \mp \right) $ signifies spin down or up.

\subsection{The Singular Even-Power potential}

In the noncommutative complex space the deformed Hamiltonian operator $\hat{H%
}_{NC}$ associated by the Singular Even-Power potential or Ivers-Power
potential is given by:

\begin{equation}
\hat{H}_{NC}=\frac{2}{m}p_{\hat{z}}p_{\widehat{\bar{z}}}+V\left( \hat{r}%
\right)   \tag{41}
\end{equation}

where $V\left( \widehat{r}\right) $ is taken to be:

\begin{equation}
V\left( \widehat{r}\right) =\left( 
\begin{array}{cc}
a\hat{z}\widehat{\bar{z}}+\frac{b}{\hat{z}\widehat{\bar{z}}}+\frac{c}{\hat{z}%
\widehat{\bar{z}}^{2}}+\frac{d}{\hat{z}\widehat{\bar{z}}^{3}} & 0 \\ 
0 & a\widehat{\bar{z}}\hat{z}+\frac{b}{\widehat{\bar{z}}\hat{z}}+\frac{c}{%
\widehat{\bar{z}}\hat{z}^{2}}+\frac{d}{\widehat{\bar{z}}\hat{z}^{3}}%
\end{array}%
\right) \text{ }  \tag{42}
\end{equation}

\bigskip For simplicity we take $\theta _{i}$ $=$ $\delta _{i3}\theta $ and
assume that the other components are all zero,then the Singular Even-Power
potential $V\left( \hat{z},\widehat{\bar{z}}\right) $ as follows:

\begin{eqnarray}
V\left( \hat{z},\widehat{\bar{z}}\right)  &=&ar^{2}+br^{-2}+\hat{c}r^{-4}+%
\hat{d}r^{-6}-\theta a\left( m\mp 1\right) +V_{NC}^{\theta \pm }\left(
r\right)   \notag \\
&=&\hat{V}_{O}^{\pm }(r)+V_{NC}^{\theta \pm }\left( r\right)   \TCItag{43}
\end{eqnarray}

where

\begin{equation}
\hat{c}=c+b\theta \left( m+2s_{z}\right) ,\text{ \ }\hat{d}=d+c\theta \left(
m+2s_{z}\right) ,\text{ }s_{z}=\mp 1/2.  \tag{44}
\end{equation}

and

\begin{equation}
\hat{V}^{\pm }(r)=ar^{2}+br^{-2}+\hat{c}r^{-4}+\hat{d}r^{-6}-\theta a\left(
m+2s_{z}\right)   \tag{45}
\end{equation}

In noncommutative complex space the eq. $(40)$ accepted a solution exact
with noncommutative Singular Even-Power potential $\hat{V}_{O}^{\pm }(r)$
for the wave functions $\Psi ^{\pm }\left( z,\bar{z}\right) $, as given by:

\begin{equation}
\Psi ^{\pm }\left( z,\bar{z}\right) =\Phi \left( \varphi \right) R_{m}\left(
z,\bar{z}\right) \left\vert \pm \right\rangle   \tag{46}
\end{equation}

Where the angular function $\Phi \left( \varphi \right) $ and radial
function $R_{m}\left( z,\bar{z}\right) $, as follows, respectively $[8,13]$:

\begin{equation}
\left\{ 
\begin{array}{c}
\Phi \left( \varphi \right) =\frac{1}{\sqrt{2\pi }}\exp \left( \pm im\varphi
\right) \text{ where }m=0;1,2... \\ 
R_{m}\left( z,\bar{z}\right) =\exp \left( \hat{p}_{m}\left( z,\bar{z}\right)
\right) \underset{n=0}{\sum }a_{n}\left( z\bar{z}\right) ^{\frac{2n+\hat{%
\delta}}{2}}%
\end{array}%
\right.   \tag{47}
\end{equation}

and $p_{m}\left( z,\bar{z}\right) $ is given by$:$

\begin{equation}
p_{m}\left( z,\bar{z}\right) =\frac{1}{2}\sqrt{a}z\bar{z}+\frac{1}{2}\sqrt{%
\hat{d}}\left( z\bar{z}\right) ^{-1}  \tag{48}
\end{equation}

For $n=0$ and $n=1$ we have, the radial functions and the energies
corresponding the stationary state and first excited states, respectively:

\begin{equation}
\left\{ 
\begin{array}{c}
\begin{array}{c}
R_{m}^{\left( 0\right) }=a_{0}\left( z\bar{z}\right) ^{\hat{\delta}/2}\exp
\left( -\frac{\sqrt{a}\left( z\bar{z}\right) ^{2}+\sqrt{\hat{d}}}{2z\bar{z}}%
\right)  \\ 
E_{0}=\sqrt{a}\left( 4+2\hat{\mu}\right) -\theta a\left( m\mp 1\right) 
\end{array}
\\ 
\text{and} \\ 
\begin{array}{c}
R_{m}^{\left( 1\right) }=\left( a_{0}+a_{1}z\bar{z}\right) \left( z\bar{z}%
\right) ^{\hat{\delta}/2}\exp \left( -\frac{\sqrt{a}\left( z\bar{z}\right)
^{2}+\sqrt{\hat{d}}}{2z\bar{z}}\right)  \\ 
E_{1}=\sqrt{a}\left( 8+2\hat{\mu}\right) -\theta a\left( m\mp 1\right) 
\end{array}%
\end{array}%
\right.   \tag{49}
\end{equation}

and the second term $V_{NC}^{\theta \pm }\left( r\right) $ in eq.$(43)$, is
the perturbation term up to the second order of $\theta $ wich takes the form%
$:$

$:$

\begin{eqnarray}
V_{NC}^{\theta \pm }\left( r\right)  &=&V_{\text{pert}}^{\theta \pm }\left(
r\right) =\theta \frac{3d}{r^{8}}\left( m\mp 1\right)   \notag \\
&=&\theta f\left( r\right) \left( m-2s_{z}\right)   \TCItag{50}
\end{eqnarray}

wher

\begin{equation}
f\left( r\right) =\frac{3d}{r^{8}}  \tag{51}
\end{equation}

and $s_{z}=\mp \frac{1}{2}$, described a particle of sipin $1/2$.

Hence we can say that the Schrdinger equation in non-commutative complex
space describes to the particles with spin ($1/2$) in an external uniform
magnitic field. To investigate the modication of the energy levels by eq. $%
(50)$, we use the rst-order perturbation theory. The spectrum of $%
ar^{2}+br^{-2}+\hat{c}r^{-4}+\hat{d}r^{-6}-\theta a\left( m+2s_{z}\right) $
and the corresponding wave functions are well-known and given by:

\begin{equation}
R_{m}=\left( a_{0}+a_{1}r^{2}+...a_{p}r^{2p}\right) r^{\hat{\delta}}\exp
\left( -\frac{\sqrt{a}}{2}r^{2}-\frac{\sqrt{\hat{d}}}{2}r^{-2}\right)  
\tag{52}
\end{equation}

where the noncommutative energy levels are given by:

\begin{equation}
\hat{E}_{n}=\sqrt{a}\left( 2\hat{\mu}+4+4n\right) -\theta a\left( m\mp
1\right)   \tag{53}
\end{equation}

where

\begin{equation}
\hat{\mu}=\frac{\hat{c}}{2\sqrt{\hat{d}}}  \tag{54}
\end{equation}

Now to obtain the the correction of the energy levels $E_{NC}^{\theta +}$\
associate withe spin up and $E_{NC}^{\theta -}$\ associate with spin down to
first order in $\theta $, for the perturbation potential $V_{NC}^{\theta \pm
}\left( r\right) $ are given by:%
\begin{eqnarray}
E_{NC}^{\theta \pm } &=&\left\langle n\right\vert V_{NC}^{\theta +}\left(
r\right) \left( r\right) \left\vert n\right\rangle =-\theta \left( m\pm
1\right) \int \Psi ^{\pm \left( p\right) \ast }\left( z,\bar{z}\right)
f\left( z,\bar{z}\right) \Psi ^{\pm \left( p\right) }\left( z,\bar{z}\right)
rdr  \notag \\
&=&-\theta \left( m\pm 1\right) \int R^{\left( p\right) \ast }\left(
r\right) f\left( r\right) R^{\left( p\right) }\left( r\right) rdr 
\TCItag{55}
\end{eqnarray}

Now to obtain the modification to the energy levels for the $n=0$ as a
result of the noncommutative term in eqs. $(50)$ , we use the first-order
perturbation theory. The expectation value of \ $r^{-8}$ with respect to the
exact solution, are given by:

\begin{equation}
\text{ }D=3da_{0}\overset{+\infty }{\underset{0}{\dint }}r^{2\hat{\delta}%
-7}\exp \left( -\sqrt{a}r^{2}-\sqrt{\hat{d}}r^{-2}\right) dr  \tag{56}
\end{equation}

we using the following standard integtal $\left[ 15\right] :$

\begin{equation}
\overset{+\infty }{\underset{0}{\dint }}r^{v-1}\exp \left( -\left( \frac{%
\lambda _{2}}{r}+\lambda _{1}r\right) \right) dr=2\left( \frac{\lambda _{2}}{%
\lambda _{1}}\right) ^{\frac{v}{2}}K_{v}\left( 2\sqrt{\lambda _{1}\lambda
_{2}}\right)   \tag{57}
\end{equation}

where $K_{v}$ the modified Bessel function of second kind and order $v$ and $%
\lambda _{1}$ and $\lambda _{2}$ are positive and $\left\vert 2\sqrt{\lambda
_{1}\lambda _{2}}\right\vert \langle \frac{\pi }{2}$ , after the explicit
calculation, the term of eq$.(56)$ take the form:

\begin{equation}
\text{ }D=\frac{3}{2}da_{0}\left( \sqrt{\frac{\hat{d}}{a}}\right) ^{\frac{%
\hat{\delta}-3}{2}}K_{\left( \hat{\delta}-3\right) }\left( 2\left( a\hat{d}%
\right) ^{\frac{1}{4}}\right)   \tag{58}
\end{equation}

Hence the modication to the energy levels $E_{0,m}^{\theta \pm }$ is given
by:

\begin{equation}
E_{0,m}^{\theta \pm }=\frac{3}{2}da_{0}\left( m\mp 1\right) \theta \left( 
\sqrt{\frac{\hat{d}}{a}}\right) ^{\frac{\hat{\delta}-3}{2}}K_{\left( \hat{%
\delta}-3\right) }\left( 2\left( a\hat{d}\right) ^{\frac{1}{4}}\right) . 
\tag{59}
\end{equation}

and the non-commutative correction of the energy levels, corresponding the
first excited states $E_{1,m}^{\theta \pm }$, in the first order of $\theta $%
:

\begin{eqnarray}
E_{1,m}^{\theta \pm } &=&\left( m\mp 1\right) \theta \overset{+\infty }{%
\underset{0}{\dint }}\left[ \left( a_{0}+a_{1}r^{2}\right) r^{\hat{\delta}-%
\frac{1}{2}}\exp \left( -\frac{\sqrt{a}r^{2}+\sqrt{\hat{d}}r^{-2}}{2}\right) %
\right] ^{2}\times   \notag \\
&&\left( \frac{3d}{r^{8}}\right) rdr  \notag \\
&=&\left( m\mp 1\right) \theta \left( \sum_{i=1}^{3}A^{i}\right)  
\TCItag{60}
\end{eqnarray}

\ where

\begin{equation}
\begin{array}{c}
A^{1}=\frac{3da_{0}^{2}}{2}\left( \sqrt{\frac{\hat{d}}{a}}\right) ^{\frac{%
\hat{\delta}-3}{2}}K_{\hat{\delta}-3}\left( 2\left( a\hat{d}\right) ^{\frac{1%
}{4}}\right) , \\ 
A^{2}=3da_{1}a_{0}\left( \sqrt{\frac{\hat{d}}{a}}\right) ^{\frac{\hat{\delta}%
-2}{2}}K_{\hat{\delta}-2}\left( 2\left( a\hat{d}\right) ^{\frac{1}{4}%
}\right) , \\ 
A^{3}=\frac{3d}{2}a_{1}^{2}\left( \sqrt{\frac{\hat{d}}{a}}\right) ^{\frac{%
\hat{\delta}-1}{2}}K_{\hat{\delta}-1}\left( 2\left( a\hat{d}\right) ^{\frac{1%
}{4}}\right) .%
\end{array}
\tag{61}
\end{equation}

know by the same method, the non-commutative modification of the energy
levels $E_{p,m}^{\theta \pm }$ to the p$^{th}$ of excitation state, up in
the first order of $\theta $ :%
\begin{eqnarray}
E_{p,m}^{\theta \pm } &=&\left( m\mp 1\right) \theta \overset{+\infty }{%
\underset{0}{\dint }}\left[ \left( a_{0}+a_{1}r^{2}+...a_{p}r^{2p}\right) %
\right] ^{2}r^{2\delta }\exp \left( -\frac{\sqrt{a}}{2}r^{2}-\frac{\sqrt{%
\hat{d}}}{2}r^{-2}\right) \times   \notag \\
&&\left( \frac{3d}{r^{8}}\right) dr  \notag \\
&=&\left( m\mp 1\right) \theta A  \TCItag{62}
\end{eqnarray}

\bigskip where

\begin{eqnarray}
A &=&\overset{+\infty }{\underset{0}{\dint }}\left[ \left(
a_{0}+a_{1}r^{2}+...a_{p}r^{2p}\right) \right] ^{2}r^{2\delta }\exp \left( -%
\frac{\sqrt{a}}{2}r^{2}-\frac{\sqrt{\hat{d}}}{2}r^{-2}\right) \times   \notag
\\
&&\left( \frac{3d}{r^{8}}\right) rdr  \TCItag{63}
\end{eqnarray}

The energy levles to the p$^{th}$ of excitation state in noncommutative
complex space as :

\begin{equation}
\hat{E}=\hat{E}_{p,m}+\left( m\mp 1\right) \theta A  \tag{64}
\end{equation}

We have shown that the non-commutativity induces the lamb shift where the
energy spectrum depends on $m$ and is splitting of two levels. Then we
deduced that the non-commutativity plays the role of magnetic field, it also
creates a spin of particle. .

\subsection{The Invers-Power potential}

In the noncommutative complex space the deformed Ivers-Power potential is
given by:

\begin{equation}
V\left( \widehat{r}\right) =\left( 
\begin{array}{cc}
\frac{a}{\sqrt{\hat{z}\widehat{\bar{z}}}}+\frac{b}{\hat{z}\widehat{\bar{z}}}+%
\frac{c}{\hat{z}\widehat{\bar{z}}\sqrt{\hat{z}\widehat{\bar{z}}}}+\frac{d}{%
\hat{z}\widehat{\bar{z}}^{2}} & 0 \\ 
0 & \frac{a}{\sqrt{\widehat{\bar{z}}\hat{z}}}+\frac{b}{\widehat{\bar{z}}\hat{%
z}}+\frac{c}{\widehat{\bar{z}}\hat{z}\sqrt{\widehat{\bar{z}}\hat{z}}}+\frac{d%
}{\widehat{\bar{z}}\hat{z}^{2}}%
\end{array}%
\right)   \tag{65}
\end{equation}

For simplicity we take $\theta _{i}$ $=$ $\delta _{i3}\theta $ and assume
that the other components are all zero,then the Ivers-Power potentiall $%
V\left( \hat{z},\widehat{\bar{z}}\right) $ as follows:

\begin{equation}
V\left( \widehat{r}\right) =ar^{-1}+br^{-2}+\hat{c}r^{-3}+\hat{d}%
r^{-4}+V_{NC}^{\theta \pm }\left( r\right)   \tag{66}
\end{equation}

where

\begin{equation}
\hat{c}=c+a\theta \left( m+2s_{z}\right) ,\text{ }\hat{d}=d-b\theta \left(
m+2s_{z}\right)   \tag{67}
\end{equation}

and the second term $V_{NC}^{\theta \pm }\left( r\right) $ is the
perturbation term up to the second order of $\theta $ wich takes the form$:$

\begin{eqnarray}
V_{NC}^{\theta \pm }\left( r\right)  &=&\theta \left( \frac{2c}{r^{5}}+\frac{%
3d}{r^{6}}\right) \left( L_{z}\mp 2s_{z}\right)   \notag \\
&=&\theta g\left( r\right) \left( L_{z}+2s_{z}\right) ,\text{ \ \ \ }%
s_{z}=\mp 1/2  \TCItag{68}
\end{eqnarray}

where

\begin{equation}
g\left( r\right) =\frac{2c}{r^{5}}+\frac{3d}{r^{6}}  \tag{69}
\end{equation}

Now the radial functions and the energies corresponding the stationary state
and first excited states, respectively withe exact solution, as follows$%
\left[ 14\right] $ :

\begin{eqnarray}
R_{0} &=&N_{0}\left( z\bar{z}\right) ^{c_{0}/2}\exp \left( \frac{a+bz\bar{z}%
}{\sqrt{z\bar{z}}}\right) ,  \TCItag{70} \\
\hat{E}_{0}^{\pm } &=&\frac{1}{16a}\left[ \hat{\lambda}_{0}\pm \sqrt{\hat{%
\lambda}_{0}^{2}-2b\hat{d}}\right] ^{2}  \TCItag{71}
\end{eqnarray}

where

\begin{equation}
\hat{\lambda}_{0}=\mu \left( 1+\mu \right) +\frac{\hat{d}\sqrt{a}}{1+\mu }%
,\mu =\frac{b}{2\sqrt{a}};c_{0}=\lambda _{0}+\frac{1}{4}  \tag{72}
\end{equation}

and

\begin{eqnarray}
R_{1}^{\left( 1\right) } &=&N_{1}\left( \sqrt{z\bar{z}}-\sigma _{1}^{\left(
1\right) }\right) z\bar{z}^{\frac{c_{1}}{2}}\exp \left( \frac{a+bz\bar{z}}{%
\sqrt{z\bar{z}}}\right) ,  \TCItag{73} \\
\hat{E}_{1}^{\pm } &=&-\left( \frac{\hat{\lambda}_{1}\pm \left[ \hat{\lambda}%
_{1}^{2}-4\hat{d}\left( \sigma _{1}^{\left( 1\right) }+\sqrt{a}\right)
\left( 1+\mu \right) \right] ^{1/2}}{4\left( \sigma _{1}^{\left( 1\right) }+%
\sqrt{a}\right) }\right) ^{2}  \TCItag{74}
\end{eqnarray}

where

\begin{equation}
c_{1}=1+\mu ;\hat{\lambda}_{1}=\left( 1+\mu \right) \left( 2+\mu \right) +%
\frac{\hat{d}\sqrt{a}}{2+\mu }\left( \sigma _{1}^{\left( 1\right) }+\sqrt{a}%
\right)   \tag{75}
\end{equation}

Then the correction of the energy levels for the Ivers-Power potentiall are
given by: 
\begin{eqnarray}
E_{NC}^{\theta \pm } &=&\left\langle n\right\vert V_{NC}^{\theta +}\left(
r\right) \left( r\right) \left\vert n\right\rangle =\theta \left( m\pm
1\right) \int \Psi ^{\pm \left( p\right) \ast }\left( z,\bar{z}\right)
g\left( z,\bar{z}\right) \Psi ^{\pm \left( p\right) }\left( z,\bar{z}\right)
rdr  \notag \\
&=&\theta \left( m\pm 1\right) \int R^{\left( p\right) \ast }\left( r\right)
g\left( r\right) R^{\left( p\right) }\left( r\right) rdr  \TCItag{76}
\end{eqnarray}

Now to obtain the modification to the energy levels for the $n=0$ as a
result of the noncommutative terms in eqs. $(69)$ , we use the first-order
perturbation theory. The expectation value of \ $r^{-5}$ and $r^{-6}$ with
respect to the exact solution, are given by:

\begin{eqnarray}
\left\langle r^{-k}\right\rangle  &=&N_{0}^{2}\int drr^{2c+1-k}\exp 2\left(
ar^{-1}+br\right)   \notag \\
&=&f\left( k\right) ,\text{ \ \ \ \ }k=5,6  \TCItag{77}
\end{eqnarray}

Putting these results together one gets the modifications of the energy:

\begin{equation}
E_{NC}^{\theta \pm }=\theta \left( m\pm 1\right) \left( f\left( 5\right)
+f\left( 6\right) \right)   \tag{78}
\end{equation}

The eigenvalues are $E_{NC}^{\theta \pm }$ ,Correspond to the energy values
for the charged particule withe spin $1/2$ in magnitic field, where the
noncommutativity play the role of magnitic field withe created the total
magnetic moment of particle with spin $1/2$ ,who in turn shifted the
spectrume of energy, therefore degeneracy is removed.

This result is important because it reects the existence of Lamb shift,
which is induced by the non-commutativity of the space. Obviously, when $%
\theta =0$; then $E_{NC}^{\theta \pm }$ $=0$, which is exactly the result of
the space-space commuting case, where the energy-levels are not shifted.

Now we can written the noncommutative Hamiltonian $\left( 38\right) $ in the
noncommutative complex space as:%
\begin{equation}
\hat{H}=\left( 
\begin{array}{cc}
\hat{H}_{\hat{z}\widehat{\bar{z}}} & 0 \\ 
0 & \hat{H}_{\widehat{\bar{z}}\hat{z}}%
\end{array}%
\right)   \tag{79}
\end{equation}

where $\hat{H}_{\hat{z}\widehat{\bar{z}}}$ and $\hat{H}_{\widehat{\bar{z}}%
\hat{z}}$ are given by:

\begin{equation}
\hat{H}_{\hat{z}\widehat{\bar{z}}}=\frac{2}{m_{0}}p_{\hat{z}}p_{\widehat{%
\bar{z}}}+V\left( r\right) -\theta f\left( r\right) \left(
L_{z}+2s_{z}\right) \text{ }\equiv \hat{H}\text{\ }^{-}\text{, where \ }%
s_{z}=-\frac{1}{2}\text{ \ \ \ }  \tag{80}
\end{equation}

and

\begin{equation}
\hat{H}_{\widehat{\bar{z}}\hat{z}}=\frac{2}{m_{0}}p_{\hat{z}}p_{\widehat{%
\bar{z}}}+V\left( r\right) -\theta h\left( r\right) \left(
L_{z}+2s_{z}\right) \equiv \hat{H}^{+}\text{,\ where \ }s_{z}=+\frac{1}{2} 
\tag{81}
\end{equation}

Then the noncommutative Hamiltonian $\left( 79\right) $, as follows:

\begin{equation}
\hat{H}=\left( 
\begin{array}{cc}
\begin{array}{c}
\frac{2}{m_{0}}p_{z}p_{\bar{z}}+V\left( r\right) - \\ 
-\theta h\left( r\right) \left( L_{z}+2s_{z}\right) =\hat{H}^{-}%
\end{array}
& 0 \\ 
0 & 
\begin{array}{c}
\frac{2}{m_{0}}p_{z}p_{\bar{z}}+V\left( r\right) - \\ 
-\theta h\left( r\right) \left( L_{z}+2s_{z}\right) =\hat{H}^{+}%
\end{array}%
\end{array}%
\right)   \tag{82}
\end{equation}

\bigskip\ The Hamiltonian in $(82)$ takes the form:

\begin{equation}
\hat{H}=H_{or}+H_{NC}^{\theta }  \tag{83}
\end{equation}

where $H_{or}$ is the ordinary Hamiltonian given by:

\begin{equation}
H_{or}=\left( \frac{2}{m_{0}}p_{z}p_{\bar{z}}+V\left( r\right) \right)
I_{2\times 2}  \tag{84}
\end{equation}

where $I_{2\times 2}$ is the identity matrix united in $2D$ space, and $%
H_{NC}^{\theta }$ is given by:

\begin{equation}
H_{NC}^{\theta }=h\left( r\right) \mathbf{g}_{j}J\cdot \theta I_{2\times 2} 
\tag{85}
\end{equation}

\bigskip where $h\left( r\right) $ is the radial function $f\left( r\right) $
or the radial function $g\left( r\right) ,$ $\mathbf{g}_{j}$ Lande factor
and $\ J=L+s$ , is the total angular momentum. Thus is similar to the Zeeman
effect, this proofs that the non-commutativity has an effect similar to the
Zeeman effects which iduced by the magnetic field $[14]$ ,where the
non-commutativity leads the role of the magnetic field. This represents Lamb
shift corrections for $l=0$. This result is very important: as a possible
means of introducing electron spin we replace $l=\pm \left( j+1/2\right) $%
and $n\rightarrow n-j-1-1/2$, where $j$ is the quantum number associated to
the total angular momentum. Then the $l=0$ state has the same total quantum
number $j=1/2$. In this case the non-commutative value of the energy levels
indicates the splitting of $1s$ states.

\section{Conclusion}

In this paper we started from quantum particule with the central potentials,
the singular even-power and Invers-Power potentials in a canonical
non-commutative coplex space, using the Moyal product method, we have
derived the defermed Schrodinger equation. Using the power series expansion
method to solving and we found that the noncommutative energy is shifted to $%
\left( 2j+1\right) $ levels ,it acts here like a Lamb shift in Dirac theory.
This proofs that the non-commutativity has an effect similar to the Zeeman
effects which iduced by the magnetic field $\left[ 16\right] $ ,where the
non-commutativity leads the role of the magnetic field and it also creates a
spin of particle.

\bigskip

\end{document}